\documentclass[useAMS,usenatbib,usegraphicx]{mn2e}

\usepackage[T1]{fontenc}
\usepackage{pslatex}
\usepackage{color}
\usepackage{amsmath}
\usepackage{url}

\def\gtrsim{\mathrel{\hbox{\rlap{\hbox{\lower3pt\hbox{$\sim$}}}\hbox{\raise2pt\hbox{$>$}}}}}

\definecolor{orange}{rgb}{1,0.3,0}
\definecolor{purple}{rgb}{1,0,1}

\newcommand{\lxt}{$L_{\rm X}-T$ }

\title[Evolution of the \lxt relation from XCS-DR1]{The \textit{XMM} Cluster Survey: Evidence for energy 
injection at high redshift from evolution of the X-ray luminosity--temperature relation}

\author[Hilton et al. (XCS collaboration)]{\parbox{\textwidth}{\raggedright
Matt~Hilton$^{1,2}$\thanks{E-mail: matthew.hilton@nottingham.ac.uk},
A.~Kathy~Romer,$^{3,4}$
Scott~T.~Kay,$^{5}$ 
Nicola~Mehrtens,$^{3}$ 
E.~J.~Lloyd-Davies,$^{3}$
Peter~A.~Thomas,$^{3}$
Chris~J.~Short,$^{3}$
Julian~A.~Mayers,$^{3}$ 
Philip~J.~Rooney,$^{3}$
John~P.~Stott,$^{6}$ 
Chris~A.~Collins,$^{7}$
Craig~D.~Harrison,$^{8}$
Ben~Hoyle,$^{9}$ 
Andrew~R.~Liddle,$^{3,4}$ 
Robert~G.~Mann,$^{10}$  
Christopher~J.~Miller,$^{8}$
Martin~Sahl{\'e}n,$^{11}$
Pedro~T. P.~Viana,$^{12,13}$
Michael~Davidson,$^{10}$
Mark~Hosmer,$^{3}$
Robert~C.~Nichol,$^{14,4}$
Kivanc~Sabirli,$^{3}$
S.~A.~Stanford$^{15,16}$ and 
Michael~J.~West$^{17}$
\\
}\vspace{0.4cm}\\
\parbox{\textwidth}{\raggedright 
$^{1}$~Centre for Astronomy \& Particle Theory, School of Physics \& Astronomy, University of Nottingham, Nottingham, NG7 2RD, UK\\
$^{2}$~Astrophysics \& Cosmology Research Unit, School of Mathematics, Statistics \& Computer Science, University of KwaZulu-Natal, Private Bag X54001, Durban, 4000, SA\\
$^{3}$~Astronomy Centre, University of Sussex, Falmer, Brighton, BN1 9QH, UK\\
$^{4}$~SEPnet, South East Physics Network (\url{http://www.sepnet.ac.uk})\\
$^{5}$~Jodrell Bank Centre for Astrophysics, School of Physics and Astronomy, The University of Manchester, Manchester, M13 9PL, UK\\
$^{6}$~Extragalactic and Cosmology Group, Department of Physics, University of Durham, South Road, Durham DH1 3LE\\
$^{7}$~Astrophysics Research Institute, Liverpool John Moores University, Twelve Quays House, Egerton Wharf, Birkenhead, CH41 1LD, UK\\
$^{8}$~Astronomy Department, University of Michigan, Ann Arbor, MI 48109, USA\\ 
$^{9}$~Institut de Ci\`{e}ncies del Cosmos (ICCUB), Departmento de F\'{\i}sica, Mart\'{\i} i Franqu\'{e}s 1, 08034 Barcelona, Spain\\ 
$^{10}$~SUPA, Institute for Astronomy, University of Edinburgh, Royal Observatory, Edinburgh, EH9 3HJ, UK\\ 
$^{11}$~The Oskar Klein Centre for Cosmoparticle Physics, Department of Physics, Stockholm University, AlbaNova, SE-106 91 Stockholm, Sweden\\
$^{12}$~Centro de Astrof\'{\i}sica da Universidade do Porto, Rua das Estrelas, 4150-762, Porto, Portugal\\ 
$^{13}$~Departamento de F\'{\i}sica e Astronomia, Faculdade de Ci\^{e}ncias, Universidade do Porto, Rua do Campo Alegre, 687, 4169-007 Porto, Portugal\\
$^{14}$~Institute of Cosmology and Gravitation, Dennis Sciama Building, Burnaby Road, Portsmouth, PO1 3FX, UK\\ 
$^{15}$~Physics Department, University of California, Davis, CA 95616, USA\\ 
$^{16}$~Institute of Geophysics and Planetary Physics, Lawrence Livermore National Laboratory, Livermore, CA 94551, USA\\
$^{17}$~ESO, Alonso de Cordova 3107, Vitacura, Santiago, Chile\\
}}

\begin{document}

\date{Draft version: \today}

\pagerange{\pageref{firstpage}--\pageref{lastpage}} \pubyear{2012}

\maketitle

\label{firstpage}

\begin{abstract}
We measure the evolution of the X-ray luminosity--temperature ($L_{\rm X}-T$) relation since $z \sim 1.5$
using a sample of 211 serendipitously detected galaxy clusters with spectroscopic redshifts drawn from 
the \textit{XMM} Cluster Survey first data release (XCS-DR1). This is the first study spanning this redshift
range using a single, large, homogeneous cluster sample. Using an orthogonal regression technique, we find no
evidence for evolution in the slope or intrinsic
scatter of the relation since $z \sim 1.5$, finding both to be consistent with previous measurements at 
$z \sim 0.1$. However, the normalisation is seen to evolve negatively with
respect to the self-similar expectation: we find $E^{-1}(z)\,L_{\rm X} = 10^{44.67 \pm 0.09} (T/5)^{3.04 \pm 0.16} (1+z)^{-1.5 \pm 0.5}$, 
which is within $2\sigma$ of the zero evolution case. We see milder, but still negative, evolution with respect
to self-similar when using a bisector
regression technique. We compare our results to numerical simulations, where we fit simulated cluster samples
using the same methods used on the XCS data. Our data
favour models in which the majority of the excess entropy required to explain the slope of the \lxt relation is 
injected at high redshift. Simulations in which AGN feedback
is implemented using prescriptions from current semi-analytic galaxy formation models predict positive 
evolution of the normalisation, and differ from our data at more than $5\sigma$. This suggests that more 
efficient feedback at high redshift may be needed in these models.
\end{abstract}


\section{Introduction}
\label{s_intro}

The evolution of the X-ray properties of galaxy clusters records both the assembly history of the
most massive gravitationally bound structures in the universe and the thermal history of the intracluster
medium (ICM). Both X-ray luminosity ($L_{\rm X}$) and temperature ($T$) correlate with cluster mass, allowing
the evolution of the cluster mass function to be measured, and constraints on cosmological parameters,
including the dark energy equation of state, to be obtained \citep[e.g.][]{Vikhlinin_2009, MantzCosmo_2010}.
To make further progress in the use of clusters as cosmological probes, it is necessary to develop our
understanding of the physical processes which determine their observable properties.

The physics that determines the properties of the ICM is more complicated than simply the action of
gravitational collapse alone, which would result in clusters being approximately self-similar, and their observable
properties obeying simple scaling relations with well understood redshift evolution. In the case of the
\lxt relation, self-similar evolution predicts $L_{\rm X} \propto T^2$
\citep{Kaiser_1986}. However, it is well established that the relation has a steeper
slope, i.e. $L_{\rm X} \propto T^{2-3}$ \citep[e.g.][]{EdgeStewart_1991, Markevitch_1998, ArnaudEvrard_1999,
Vikhlinin_2002, Maughan_2006, Pacaud_2007, Pratt_2009, Takey_2011}. This indicates an additional source of
energy is heating the ICM, that is more effective in low mass systems. While some energy is injected by supernovae 
(SNe) within galaxies, it is likely that the
bulk of the energy comes from Active Galactic Nuclei (AGN) in the centres of clusters, as observations of low
redshift clusters show that AGN jets, seen in radio imaging, carve out
cavities in the hot gas observed at X-ray wavelengths \citep[e.g.][]{Birzan_2004, McNamara_2005,
Blanton_2011}.

Numerical simulations which include additional energy injection into the ICM, such as from AGN feedback, are
able to reproduce the observed \lxt relation at low redshift. However, different energy injection
models, which give consistent results at low redshift, give different predictions for the evolution of the
normalisation of the \lxt relation with redshift \citep[e.g.][]{Muanwong_2006, Short_2010, McCarthy_2011}. By
measuring the evolution of the \lxt relation to high redshift, constraints on these models can be
obtained. This also feeds naturally into models of galaxy formation, which invoke AGN feedback to prevent
overcooling in massive haloes: a consistent model of AGN feedback should be able to reproduce the observed
\lxt relation as well as the galaxy luminosity function \citep[e.g.][]{Bower_2008}. However, to date
there is no consensus on the evolution of the relation to high redshift: some studies find that the evolution
is consistent with self-similar \citep[e.g.][]{Vikhlinin_2002, Lumb_2004, Maughan_2006, Pacaud_2007}, while 
other studies find evidence for either zero or negative evolution \citep[e.g.][]{Ettori_2004, Reichert_2011,
Clerc_2012}. The X-ray cluster samples on which these works are based contain few clusters at high redshift, or are
heterogeneous (i.e. containing objects drawn from many different surveys), making it difficult to account for
selection effects, which can mimic evolution \citep[e.g.][]{Pacaud_2007, Short_2010}.

In this paper, we examine the evolution of the \lxt relation over the last $\sim 9$~Gyr using the
\textit{XMM} Cluster Survey \citep[XCS\footnotemark;][]{Romer_2001}. XCS is a serendipitous search for galaxy clusters in
the \textit{XMM-Newton} Science Archive. The X-ray analysis methodology for the survey is described in
\citet{LloydDavies_2011}. The first data release \citep[XCS-DR1;][]{Mehrtens_2011} contains a total of 401
X-ray selected clusters with temperature and redshift estimates, the largest such sample to date. The
sensitivity of \textit{XMM-Newton} allows XCS to detect a larger number of clusters at high redshift compared
to earlier serendipitous cluster searches conducted with \textit{ROSAT}; the XCS-DR1 catalogue
contains 38 clusters at $z > 0.5$ with spectroscopic redshifts and temperature measurements. The most distant cluster in the sample is
J2215.9-1738 at $z=1.46$ \citep{Stanford_2006, Hilton_2007, Hilton_2009, Hilton_2010}. In this work we use
this wide redshift range to measure the evolution of the \lxt relation, and therefore constrain models for 
energy injection into the ICM, such as AGN feedback \citep[e.g.][]{Short_2010}. \citet{Stott_2012} present a
complementary study of the effect of AGN feedback in groups and clusters using a low redshift ($z < 0.3$)
subsample of XCS-DR1 clusters cross matched with the FIRST catalogue of radio sources \citep{White_1997}. 
Other analyses based on XCS-DR1 include a study of fossil groups and clusters \citep{Harrison_2012}, and
\citet{Viana_2012} describes the predicted overlap with the \textit{Planck} Sunyaev-Zel'dovich effect selected
cluster catalogue \citep[][]{PlanckESZ_2011}.
\footnotetext{\url{http://www.xcs-home.org}}

The structure of this paper is as follows. In Section~\ref{s_data}, we provide a brief introduction to the
XCS-DR1 cluster sample used in this work. We describe the method used to measure the
\lxt relation and its evolution in Section~\ref{s_analysis}, and present our results in
Section~\ref{s_results}. We discuss our findings in the context of numerical simulations in 
Section~\ref{s_discussion}, and present our conclusions in Section~\ref{s_conclusions}.

We assume a cosmology of $\Omega_{\rm m}=0.27$, $\Omega_\Lambda=0.73$, and $H_0=70$ km s$^{-1}$ Mpc$^{-1}$
throughout.

\section{Data}
\label{s_data}

The XCS-DR1 cluster catalogue is presented in \citet[][]{Mehrtens_2011}, while the algorithms
used in generating the catalogue are described in detail in \citet[][LD11 hereafter]{LloydDavies_2011}, and so
here we provide only a brief summary of the data used in this paper.

\defcitealias{LloydDavies_2011}{LD11}

XCS-DR1 is constructed from 5776 \textit{XMM} observations, publicly available before July 2010. A total of 3675
extended X-ray sources (i.e. cluster candidates) were detected at $> 4 \sigma$ significance with $> 50$ counts
using a wavelet based detection algorithm, in an area covering $\sim
410$~deg$^2$ (see LD11). The majority of these cluster candidates have yet to be optically confirmed; the XCS-DR1
catalogue consists of the first batch of 401 clusters with redshift and temperature measurements 
\citep[see][]{Mehrtens_2011}.

X-ray luminosities and temperatures were measured for each cluster in XCS-DR1 using fully automated
pipelines. The temperature measurements are described in Section 4.2 of LD11. Four different
models, including one simulating the effect of undetected AGN contamination, and another simulating the effect
of a cool core, were fitted to the spectral data using \texttt{XSPEC} \citep{Arnaud_1996}, with the best
fitting model being adopted for the temperature measurement. X-ray luminosities were measured within $R_{500}$
(i.e. the radius at which the enclosed mean density is 500 times the critical density at the cluster redshift), as described
in Section 4.3 of LD11, by fitting the surface brightness profile using a $\beta$ model \citep{Cavaliere_1976}
and extrapolating to $R_{500}$ where necessary. It is important to note that unlike dedicated follow-up
observations of known clusters \citep[e.g.][]{Vikhlinin_2006, Pratt_2009, Maughan_2012}, the serendipitous data analysed by XCS
is not of sufficient quality (i.e. low counts, low resolution due to detection off-axis) to excise emission
from cluster cores. 

\begin{figure}
\includegraphics[width=8.3cm]{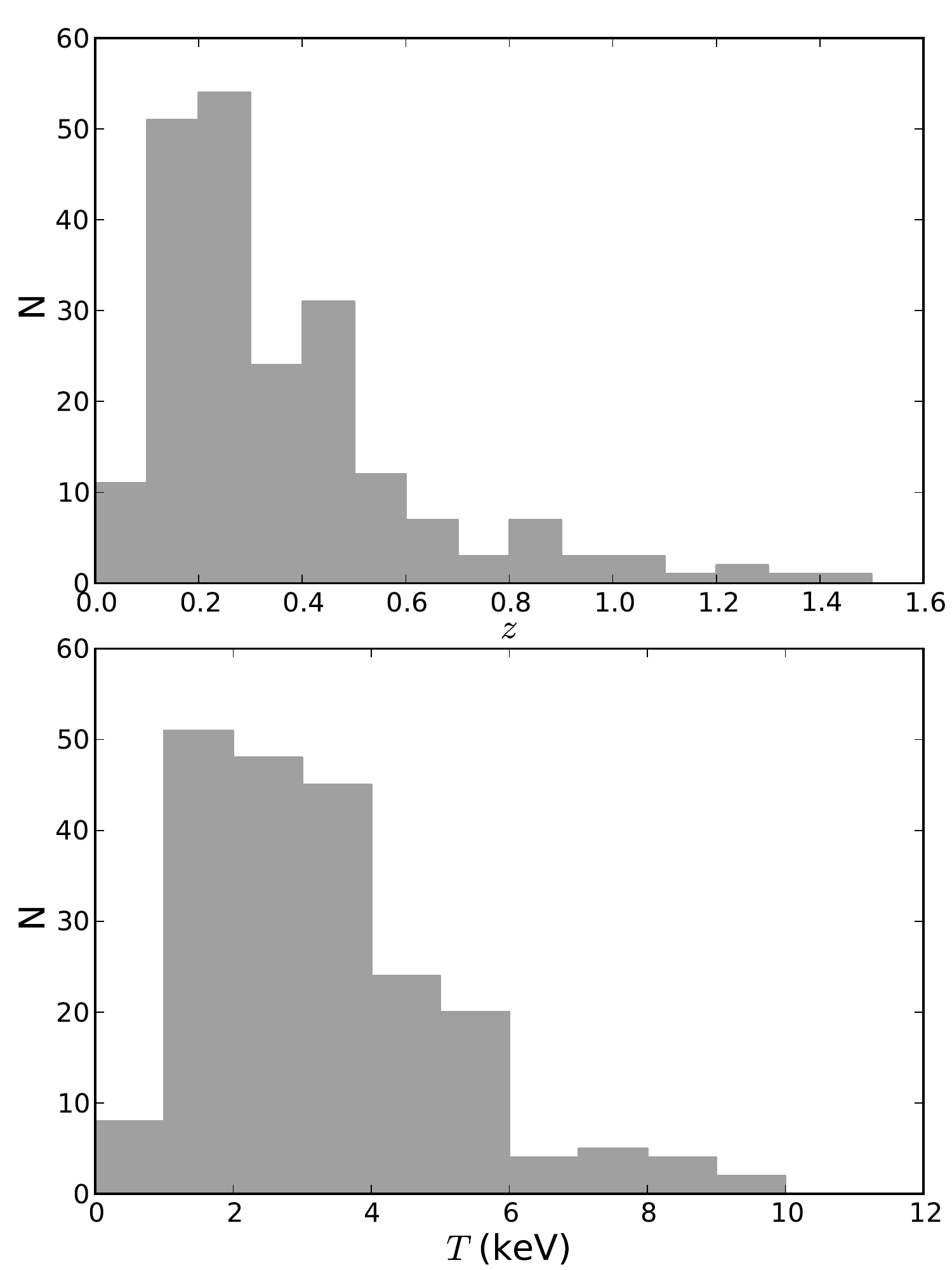}
\caption{Redshift and temperature distributions for the sample of 211 XCS-DR1 clusters with spectroscopic 
redshifts used in this work \citep[see][for a description of the catalogue]{Mehrtens_2011}.}
\label{f_zT_histograms}
\end{figure}

In this work, we use only the subsample of 211 XCS-DR1 clusters with spectroscopic redshifts. While all of the
data used in this paper is publicly available in the form of the XCS-DR1 catalogue\footnotemark, for 
completeness, the sample used here is listed in a supplementary table available with the 
online edition of this article. Fig.~\ref{f_zT_histograms} shows the redshift and temperature distributions of
the sample. The clusters span the redshift range 0.06--1.46 (median $z = 0.28$), and temperature range 0.6--9.8\,keV 
(median $T = 2.9$\,keV). Note that in the analysis presented in 
this paper, we do not attempt to correct for selection effects - given the redshift incompleteness of XCS (i.e., many candidate 
clusters within the survey area from which XCS-DR1 is drawn do not have optical follow-up or redshifts),
accounting for selection biases is not straightforward, and is deferred to future work. We do however 
comment on the expected effect of Malmquist bias on our results for the \lxt relation evolution in 
Section~\ref{s_selection}.
\footnotetext{\url{http://www.xcs-home.org/datareleases}}

\section{Analysis}
\label{s_analysis}

The large size of the XCS-DR1 catalogue allows us to simultaneously fit for the redshift evolution of the \lxt
relation, in addition to its slope, normalisation and intrinsic scatter, using a model of the form
\begin{equation}
 \log (E^{-1}(z)\,L_{\rm X}) = A + B\log (T/5) + C\log (1+z),
\label{e_model}
\end{equation}
where $L_{\rm X}$ is the bolometric X-ray luminosity measured within $R_{500}$ in erg s$^{-1}$ and $T$ is the X-ray
temperature in keV. The advantage of this approach is that it avoids the need to bin the data by redshift.
We set the pivot temperature to 5\,keV for ease of comparison with other works 
\citep[e.g.][]{Pratt_2009}, although this is higher than the median temperature ($T=2.9$\,keV) of the sample.
This model assumes that the slope of the relation does not evolve with redshift. Note that
we have scaled the luminosities by $E^{-1}(z)$ (the evolution of the
Hubble parameter, i.e. $E(z) = [\Omega_{\rm m}(1+z)^3 + \Omega_{\Lambda}]^{1/2}$), which is the evolution expected in the self-similar case, in which clusters are expected to
become more luminous at fixed temperature as redshift increases. Hence $C = 0$ corresponds to 
self-similar evolution, while $C < 0$ indicates evolution which is slower than self-similar. 

We estimate the parameters of this model using Markov Chain Monte Carlo (MCMC), using two different methods
which both take into account the intrinsic scatter and the measurement errors. Our approach is similar to
that of \citeauthor[][]{Weiner_2006}~(2006; see also \citealt{Kelly_2007}). Firstly, we define an orthogonal
regression method, for which the probability density for a given cluster to be drawn from this model is
\begin{equation}
 P_{\rm model} = \cfrac{1}{\sqrt{2\pi (\Delta r^2 + S^2)}} \exp \left[- \cfrac{(r-r_{\rm
model})^2}{2(\Delta r^2 + S^2)} \right],
\label{e_Pmodel}
\end{equation}
where $r-r_{\rm model}$ is the orthogonal distance of the cluster from the model relation in the $\log~L_{\rm X}$--$\log~T$
plane; $\Delta r$ is the error on the orthogonal distance, obtained from the projection in the direction
orthogonal to the model line of the ellipse defined by the errors on $\log~L_{\rm X}$, $\log~T$ (appropriate
sides of asymmetric error bars are chosen here according to the position of a given point relative to the model fit
line); and $S$ is the (orthogonal) intrinsic scatter. The latter can be converted into the scatter in 
the $\log L_{\rm X}$ axis ($\sigma_{\log L_{\rm X}}$) using
\begin{equation}
\sigma_{\log L_{\rm X}} = S/\cos(\tan^{-1} B).
\end{equation}

\begin{table}
\centering
\caption{Priors on \lxt relation fit parameters (see Section~\ref{s_analysis}).}
\label{t_priors}
\begin{tabular}{|c|c|c|}
\hline
Parameter                   & Uniform Prior & Notes\\
\hline
$A$                         & (41, 47)    & -\\
$B$                         & (1, 5)      & -\\
$C$                         & (-3, 3)     & -\\
$S$                         & (0.01, 0.5) & Orthogonal method only\\
$\sigma_{\log L_{\rm X}}$   & (0.01, 0.5) & Bisector method only\\
$\sigma_{\log T}$           & (0.01, 0.5) & Bisector method only\\
\hline
\end{tabular}
\end{table}

\begin{figure*}
\includegraphics[width=17.8cm]{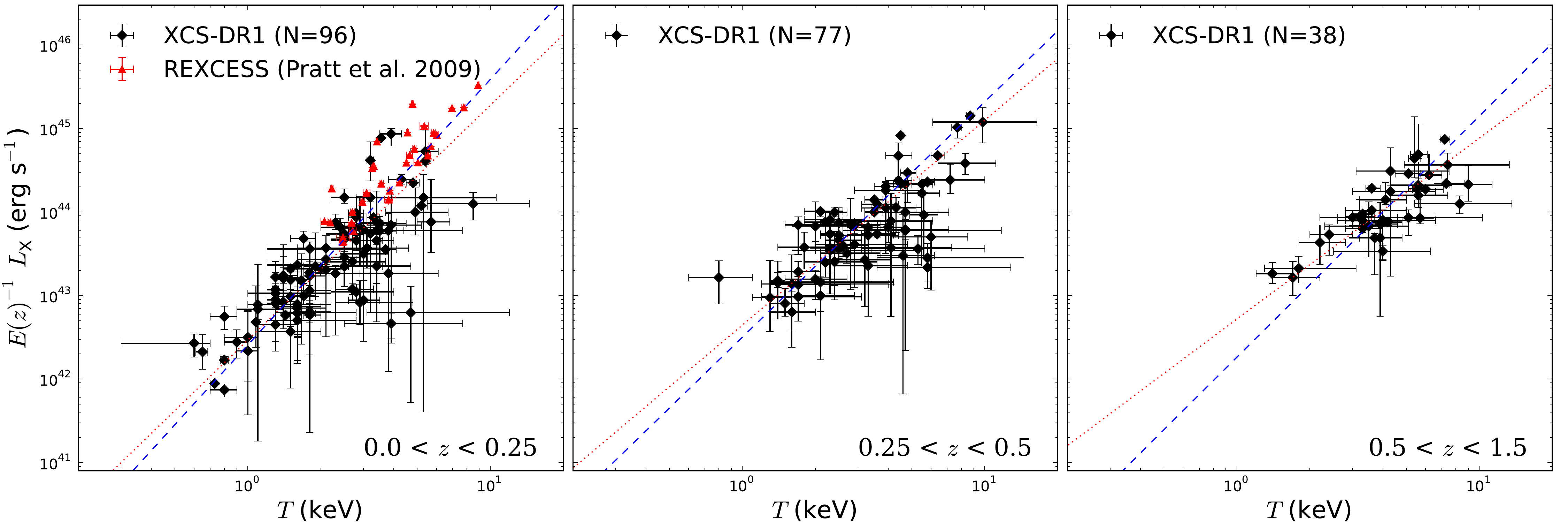}
\caption{\lxt relations for subsamples of XCS-DR1 in redshift bins. The dashed line shows the best fit to the data
from the orthogonal method, while the dotted line shows the best fit according to the bisector method (see 
Section~\ref{s_analysis}). The red triangles in the left hand panel show the REXCESS sample \citep{Pratt_2009} 
for comparison. Neither the slope nor scatter change significantly with redshift using the 
orthogonal method, while we see shallower slopes at high redshift using the bisector method (see Table~\ref{t_zbinFits}).}
\label{f_LT_sideBySide}
\end{figure*}

\begin{table*}
\caption{\lxt relation fit parameters, derived from the bisector and orthogonal methods (see Section~\ref{s_analysis}),
for XCS-DR1 subsamples in redshift bins. The model fitted is $\log (E^{-1}(z)\,L_{\rm X}) = A + B\log (T/5)$, and the units of 
$T$ and $L_{\rm X}$ are keV and erg s$^{-1}$ respectively. The uncertainties are the marginalised 68 per cent confidence regions
on each parameter derived using MCMC.}
\label{t_zbinFits}
\begin{tabular}{|l|c|c|c|c|c|c|c|}
\hline
                  & & \multicolumn{3}{|c|}{Bisector} & \multicolumn{3}{|c|}{Orthogonal}\\
Redshift range & $N$ & $A$ & $B$ & $\sigma_{\log L_{\rm X}}$ & $A$ & $B$ & $\sigma_{\log L_{\rm X}}$ \\
\hline
\phantom{0}$0.0 < z < 0.25$ & 96 & $44.43 \pm 0.06$ & $2.81 \pm 0.14$ & $0.40 \pm 0.04$ &$44.63 \pm 0.10$ & $3.18 \pm 0.22$ & $0.33 \pm 0.04$ \\
$0.25 < z < 0.5$ & 77 & $44.36 \pm 0.04$ & $2.45 \pm 0.14$ & $0.33 \pm 0.04$ &$44.47 \pm 0.07$ & $2.82 \pm 0.25$ & $0.23 \pm 0.04$ \\
\phantom{0}$0.5 < z < 1.5$ & 38 & $44.23 \pm 0.04$ & $2.17 \pm 0.19$ & $0.24 \pm 0.03$ &$44.28 \pm 0.07$ & $2.89 \pm 0.45$ & $0.24 \pm 0.05$ \\
\hline
\end{tabular}
\end{table*}

We also use a bisector method, in which the scatter and measurement errors in each axis are 
treated independently. In this case, $P_{\rm model}$ is the product of the Gaussian 
probabilities of the residuals of $L_{\rm X}$ and $T$ from the given bisector best-fitting line defined by the model 
parameters, i.e., we substitute
\begin{equation}
y_{\rm model} =\log (E^{-1}(z)\,L_{\rm X})-[A+B\log(T/5)+C\log(1+z)], 
\end{equation}
\begin{equation}
x_{\rm model} =\log(T/5)-[\log (E^{-1}(z)\,L_{\rm X})-A-C\log(1+z)]/B,
\end{equation}
instead of $r_{\rm model}$ in Equation~\ref{e_Pmodel}, and replace $r$, $\Delta r$ as appropriate. We replace
$S$ with two parameters, $\sigma_{\log L_{\rm X}}$ and $\sigma_{\log T}$.

For both methods, the likelihood $\mathcal{L}$ of a given model is simply the product of $P_{\rm model}$ for each
cluster in the sample, i.e., in the orthogonal case
\begin{equation}
\mathcal{L}(L_{\rm X}, T | A, B, C, S) \propto P_{\rm prior}(A, B, C, S) \prod_i{P_{\rm model, \emph i}},
\label{e_likelihood}
\end{equation}
where we assume generous, uniform priors on each parameter, which are listed in 
Table~\ref{t_priors}. We obtain estimates of the model parameters from the posterior distributions using MCMC,
implemented using the \citet{Metropolis_1953} algorithm. 

As shown in the next section, for $C = 0$, the results given by the bisector and orthogonal methods are 
bracketed by those obtained when using the \citet{Kelly_2007} method, with $T$ alternately used as the 
dependent or independent variable. It is important to note that there is no single method which gives the 
`true' underlying slope and normalisation for problems with errors in both variables and intrinsic scatter: 
each method gives a slope and normalisation which depends upon the assumptions in the method. Throughout this
paper, we show the results from both methods, to give an idea of the possible systematic error arising from 
the choice of fitting method.

\section{Results}
\label{s_results}

\subsection{Evolution of the slope and intrinsic scatter}
\label{s_redshiftBins}

The model for the evolution of the \lxt relation defined in Equation~\ref{e_model} assumes that there is no
evolution in the slope of the relation. We checked for this by fitting the \lxt relation of subsamples 
divided into redshift bins ($0.0 < z < 0.25$, $0.25 < z < 0.5$, and $0.5 < z < 1.5$), setting $C = 0$ in 
Equation~\ref{e_model}. Fig.~\ref{f_LT_sideBySide} and Table~\ref{t_zbinFits} show the results. 

For the lowest redshift bin ($z < 0.25$), we find a similar slope using both the orthogonal and bisector 
methods to that found in many previous studies. The values we derive are consistent with the value of 
$3.35 \pm 0.32$ measured by REXCESS\footnotemark\, \citep{Pratt_2009} at 
$z = 0.1$, as well as numerous other works \citep[e.g.][]{Markevitch_1998, ArnaudEvrard_1999, Wu_1999}, and 
most of the $z < 0.3$ subsamples of XCS-DR1 clusters in the study by \citet[][]{Stott_2012}, in which a different
fitting technique was used.
\footnotetext{We compare to REXCESS measurements with the core emission included (i.e., the $L_1$, $T_1$ values in Table~2 of \citealt{Pratt_2009}).}

\begin{figure*}
\includegraphics[width=9cm]{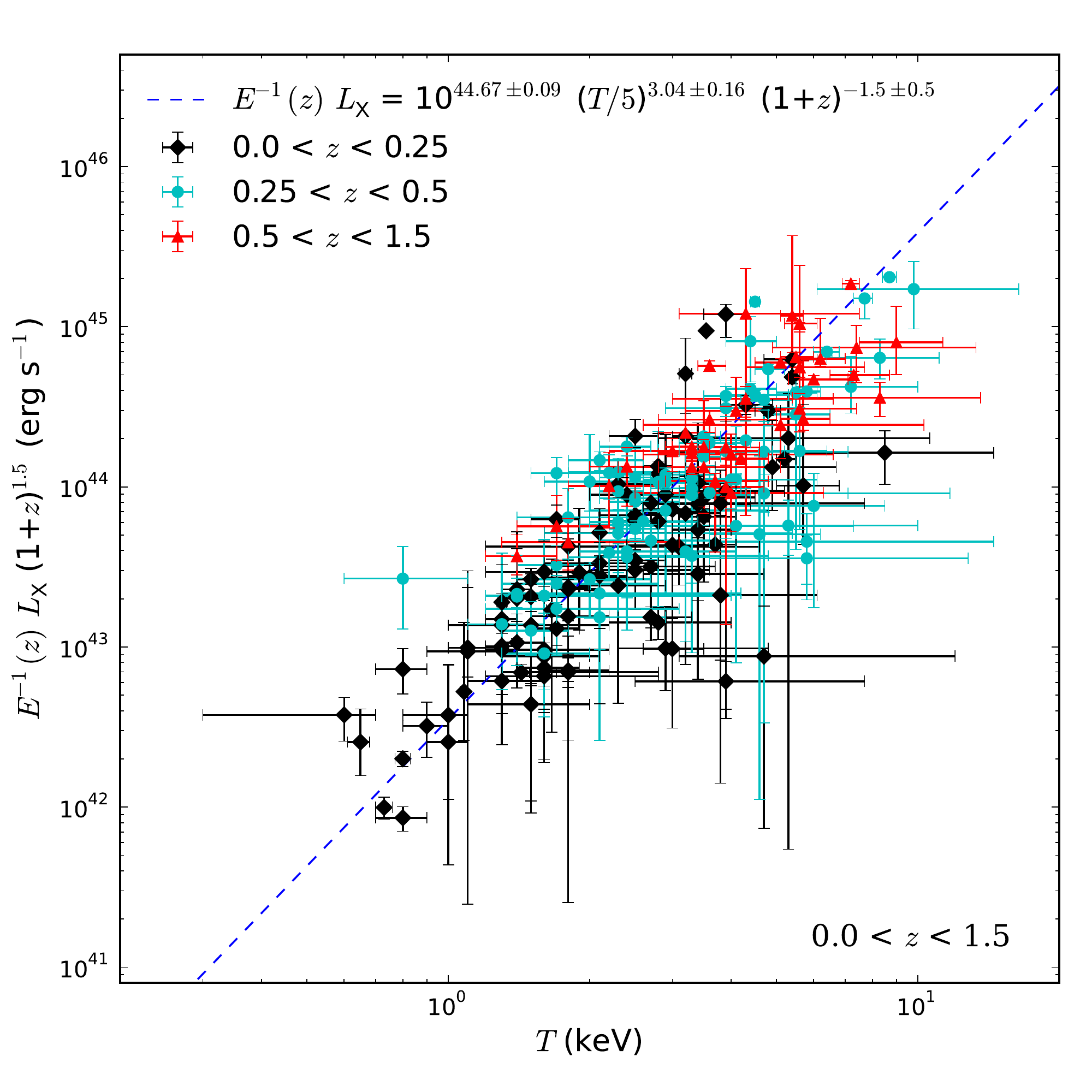}
\caption{The \lxt relation for the 211 XCS-DR1 clusters with spectroscopic redshifts. The dashed
line is the best-fitting four parameter model (Equation~\ref{e_model}), determined using the orthogonal fitting method.
The luminosities have been scaled to take into account the evolution in the normalisation as a function of 
redshift inferred from the best-fitting model parameters, 
as well as the $E^{-1}(z)$ evolution expected in the self-similar case.}
\label{f_LT_ABC}
\end{figure*}

\begin{figure*}
\includegraphics[width=11cm]{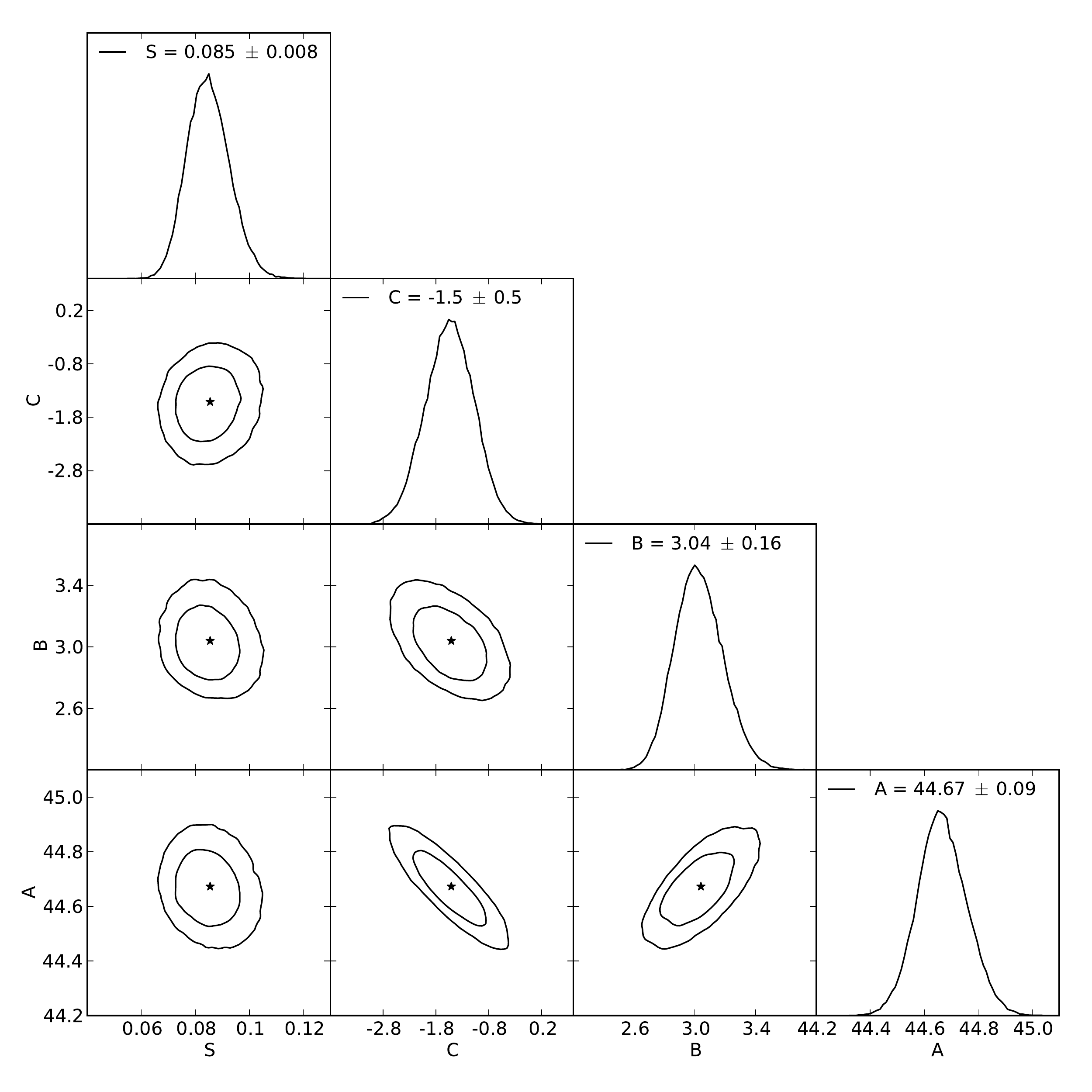}
\caption{One and two dimensional marginalised distributions (contours mark 68 and 95 per cent confidence
limits) for each combination of parameters in the four parameter evolving \lxt relation model (Equation~\ref{e_model}),
as determined using the orthogonal fitting method. Note that the luminosities have been scaled by $E^{-1}(z)$,
and so $C = 0$ corresponds to the case of self-similar evolution.}
\label{f_2DProb}
\end{figure*}

We find different values for the normalisation, depending on the fitting technique
employed. For the orthogonal method, we find $A = 44.63 \pm 0.10$, which is slightly lower, but within 2$\sigma$,
of the REXCESS value \citep[$44.85 \pm 0.06$;][]{Pratt_2009}. The normalisation obtained
using the bisector method ($A = 44.43 \pm 0.06$) is about 5$\sigma$ lower than the REXCESS value. This seems 
to be driven by the degeneracy between the slope and normalisation, with the orthogonal method 
preferring steeper slopes. As can be seen in the left hand panel Fig.~\ref{f_LT_sideBySide}, the bisector 
method gives more weight to a population of low $L_{\rm X}$, but relatively high $T$ objects, resulting in a
shallower slope and correspondingly lower normalisation. 

Clearly, as shown in the left panel of Fig.~\ref{f_LT_sideBySide}, there is not much overlap between the XCS and REXCESS temperature ranges, so it
is not surprising that there is some difference between the normalisations of the two samples. This may also
be in part due to the use of the \citet{Cash_1979} rather than $\chi^2$ statistic in the XCS spectral 
fitting (LD11; see also \citealt{Humphrey_2009}), or reflect differences in the sample selection, if for
example the XCS sample contains a smaller fraction of cool core clusters, which prefer a higher 
normalisation (see, e.g., Table~2 of \citet{Pratt_2009}, where $A = 45.11 \pm 0.16$ for cool core clusters, 
while $A = 44.70 \pm 0.03$ for non-cool core clusters).

We find that both the bisector and orthogonal fit results are bracketed by those obtained using the 
\citet{Kelly_2007} method, depending on the choice of the dependent variable. With $T$ as the independent 
variable, we find $A = 44.42 \pm 0.09$ and $B = 2.67 \pm 0.19$, which is in good agreement with our bisector 
method, although with shallower slope. For $T$ as the dependent variable, we infer $A = 44.70 \pm 0.10$ and 
$B = 3.38 \pm 0.23$ using the \citet{Kelly_2007} method, which are in good agreement with our orthogonal method, 
although with slightly higher slope and normalisation. 

The redshift evolution of the slope is different for the two fitting methods. For the orthogonal method, the slope
does not change significantly with redshift, with the values found for each subsample
differing by about $1\sigma$. This is consistent with the findings of previous studies, which have measured the slope
at $z > 0.4$ using smaller samples than that used in this work \citep[e.g.][]{Vikhlinin_2002, Novicki_2002, 
Ettori_2004, Maughan_2006, Maughan_2012, Takey_2011}. However, we do see flattening of the slope with increasing redshift 
in the fits using the bisector method. While this may be real, it is also an expected signature of Malmquist bias, 
which is discussed in Section~\ref{s_selection}.

The intrinsic scatter in the relation appears to decrease slightly with redshift (see Table~\ref{t_zbinFits}),
although the difference in the scatter between any two redshift bins (using either fitting method) is generally less than
$2\sigma$. This suggests there might be a decreasing fraction of cool core clusters at high redshift, although of course
better data are needed to determine if this is the case. We note that a decrease in the scatter at high redshift
could alternatively be due to selection effects, as shown by \citet{Reichert_2011} using simulated cluster samples. Our
intrinsic scatter estimates for the lowest redshift bin are consistent with the REXCESS measurement at $z = 0.1$ 
\citep{Pratt_2009}.

We conclude, on the basis of these results, that it is reasonable to use a model with fixed slope and scatter
to measure the evolution of the normalisation of the relation with redshift (which appears to evolve negatively
in Fig.~\ref{f_LT_sideBySide}), when using the orthogonal fitting method. We also present the results obtained
for the bisector method throughout, as this gives an indication of the possible systematic error due
to the choice of fitting technique. 

\begin{table*}
\caption{\lxt relation fit parameters, derived from the bisector and orthogonal methods (see Section~\ref{s_analysis}),
for the full XCS-DR1 cluster sample with spectroscopic redshifts. In all cases, the normalisation ($A$) is quoted at $T=5$\,keV. 
The fits with $B$ fixed have the slope set to the value found for the appropriate $0.0 < z < 0.25$ sample and fitting method 
combination listed in Table~\ref{t_zbinFits}. In all cases, the value of $C$ implies the evolution of the relation 
is below the self-similar expectation.}
\label{t_fullFits}
\begin{tabular}{|r|c|c|c|c|c|c|c|c|}
\hline
                    & \multicolumn{4}{|c|}{Bisector} & \multicolumn{4}{|c|}{Orthogonal}\\
Model & $A$ & $B$ & $C$ & $\sigma_{\log L_{\rm X}}$ & $A$ & $B$ & $C$ & $\sigma_{\log L_{\rm X}}$ \\
\hline
\multicolumn{9}{|l|}{$B$ free:}\\
$E^{-1}(z)\,L_{\rm X} \propto (1+z)^C$ & $44.41 \pm 0.05$ & $2.64 \pm 0.09$ & $-0.5 \pm 0.3$ & $0.35 \pm 0.02$ &$44.67 \pm 0.09$ & $3.04 \pm 0.16$ & $-1.5 \pm 0.5$ & $0.27 \pm 0.03$ \\
$L_{\rm X} \propto (1+z)^C$            & $44.38 \pm 0.05$ & $2.63 \pm 0.09$ & $\phantom{+}0.3 \pm 0.3$ & $0.34 \pm 0.02$ &$44.65 \pm 0.09$ & $3.03 \pm 0.16$ & $-0.7 \pm 0.5$ & $0.27 \pm 0.03$ \\
$L_{\rm X} \propto E(z)^C$             & $44.41 \pm 0.04$ & $2.65 \pm 0.09$ & $\phantom{+}0.3 \pm 0.3$ & $0.34 \pm 0.02$ &$44.63 \pm 0.07$ & $3.02 \pm 0.15$ & $-0.9 \pm 0.5$ & $0.27 \pm 0.03$ \\

\\
\multicolumn{9}{|l|}{$B$ fixed:}\\
$E^{-1}(z)\,L_{\rm X} \propto (1+z)^C$ & $44.48 \pm 0.04$ & $2.81$ & $-0.7 \pm 0.3$ & $0.36 \pm 0.02$ &$44.73 \pm 0.07$ & $3.18$ & $-1.7 \pm 0.4$ & $0.28 \pm 0.03$ \\
$L_{\rm X} \propto (1+z)^C$            & $44.46 \pm 0.04$ & $2.81$ & $\phantom{+}0.1 \pm 0.3$ & $0.36 \pm 0.02$ &$44.70 \pm 0.07$ & $3.18$ & $-0.9 \pm 0.4$ & $0.28 \pm 0.03$ \\
$L_{\rm X} \propto E(z)^C$             & $44.47 \pm 0.03$ & $2.81$ & $\phantom{+}0.0 \pm 0.3$ & $0.36 \pm 0.02$ &$44.68 \pm 0.06$ & $3.18$ & $-1.2 \pm 0.5$ & $0.28 \pm 0.03$ \\
\hline
\end{tabular}
\end{table*}

\subsection{Evolution of the normalisation}
\label{s_fullModel}
We measure the evolution of the normalisation of the $L_{\rm X}-T$ relation by fitting the four parameter 
model described by Equation~\ref{e_model} to the complete sample of 211 clusters, with $C$ now allowed to 
vary. This is the first such measurement over this redshift range using clusters drawn from a single survey
and analysed in a consistent way. Our large dataset allows us to fit for all parameters simultaneously, 
without fixing the normalisation to that measured in a different low redshift sample \citep[e.g.][]{Markevitch_1998, 
ArnaudEvrard_1999}, as has often been done in past studies of this type \citep[e.g.][]{Ettori_2004, 
Maughan_2006, Pacaud_2007}.
  
Since the bisector method shows a preference for shallower slopes at higher redshift, we focus first on the results
obtained using the orthogonal fitting method. Fig.~\ref{f_LT_ABC} presents the \lxt relation for the whole sample, with $E^{-1}(z)\,L_{\rm X}$ scaled 
according to the redshift evolution inferred from the best-fitting model
\begin{align}
 \log (E^{-1}(z)\, L_{\rm X}) = (44.67 \pm 0.09)~&+~(3.04 \pm 0.16)~\log (T/5) \nonumber \\
&-~(1.5 \pm 0.5)~\log (1+z),
\label{e_bestFit}
\end{align}
with $S = 0.085 \pm 0.008$ (i.e., $\sigma_{\log L_{\rm X}} = 0.27 \pm 0.03$). Fig.~\ref{f_2DProb} shows 
the one and two dimensional marginalised probability distributions for each parameter. We see, as expected 
given the model definition, that the slope, normalisation, and redshift evolution are all degenerate to some
extent. 

\begin{figure}
\includegraphics[width=8.3cm]{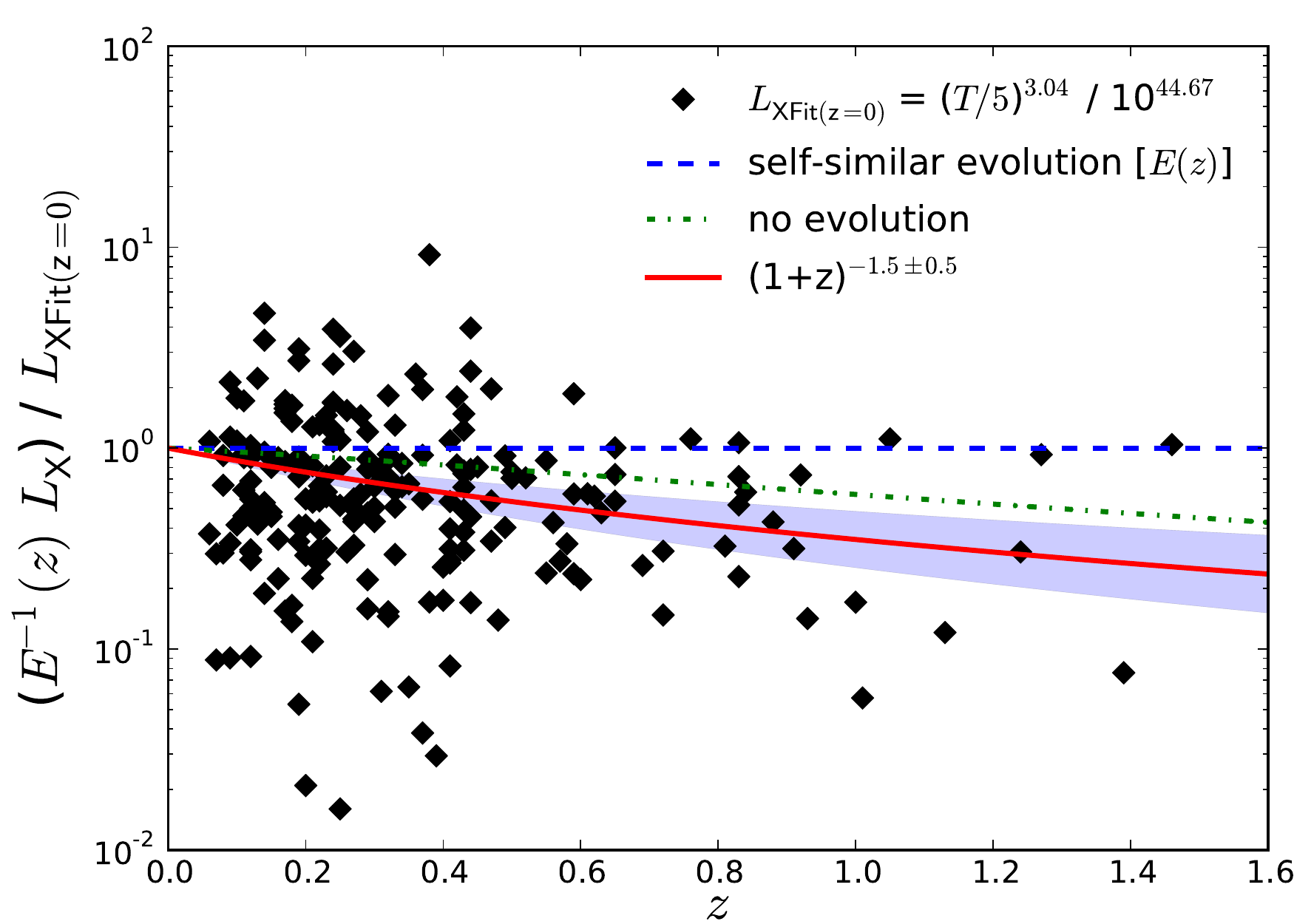}
\caption{Evolution of the normalisation of the \lxt relation relative to the self-similar case [$E(z)$], as 
inferred from the best-fitting four parameter model (Equation~\ref{e_bestFit}), using the orthogonal
fitting method. The shaded area shows the 
marginalised 68 per cent confidence region on the evolution derived using MCMC. The dot-dashed line shows the
track for no redshift evolution in the normalisation of the relation. The black diamonds show individual XCS
clusters (error bars are omitted for clarity).}
\label{f_LTNormEvo}
\end{figure}

As for the fits to the subsamples in redshift bins (Section~\ref{s_redshiftBins}), the slope and 
scatter are consistent with low redshift samples. The $z = 0$ normalisation inferred from the model 
($A = 44.67 \pm 0.09$) is slightly lower than that found in REXCESS \citep[$44.85 \pm 0.06$;][]{Pratt_2009},
but is consistent within less than $2\sigma$.

We find that the redshift evolution of the normalisation is negative ($C = -1.5 \pm 0.5$), indicating that the
evolution in luminosity at fixed temperature is significantly less than the self-similar prediction ($C = 0$).
However, the evolution we see is within $2 \sigma$ of the no evolution case. This is shown graphically in Fig.~\ref{f_LTNormEvo}. 
We checked the sensitivity of this result to reducing the redshift range, using a 
subsample of 183 clusters restricted to $z < 0.6$. We find consistent results for all parameters, although the
deviation of the normalisation from the self-similar prediction is not significant in this case ($C = -0.7 \pm 0.9$),
and is also consistent with null evolution to within less than $1\sigma$.

In Fig.~\ref{f_LTNormEvo}, we see that the highest redshift cluster in our sample, J2215.9-1738 at $z = 1.46$,
has properties consistent with self-similar evolution. This is in contrast to our previous analysis of this 
cluster \citep{Hilton_2010}, where we found it to be underluminous given its temperature. This is due
to the assumption of the \citet{Markevitch_1998} \lxt relation parameters in \citet{Hilton_2010} when 
estimating the deviation of J2215.9-1738 from self-similarity. If we adopt the $L_{\rm X}$ and $T$ 
measurements from \citet{Hilton_2010} for this system, and apply the best-fit \lxt relation parameters 
derived in this work using the orthogonal MCMC method (i.e. $A = 44.67 \pm 0.09$, $B = 3.04 \pm 0.16$), 
then we find J2215.9-1738 is well within 1$\sigma$ of the self-similar prediction.

Repeating the analysis on the whole sample using the bisector method, we find that the redshift evolution ($C = -0.5 \pm 0.3$)
is closer to self-similar than we found using the orthogonal method. The milder evolution seen in this case seems to be driven by the much lower $z = 0$ normalisation
found using the bisector method ($A = 44.41 \pm 0.05$); this is significantly (approximately 5$\sigma$) lower than the REXCESS
normalisation (see Section~\ref{s_redshiftBins} above). Since the values of the slope and normalisation are degenerate, 
and we see from the results of Section~\ref{s_redshiftBins} that the bisector method favours shallower slopes at high redshift, we repeated the fit
with the value of the slope fixed to $B = 2.81$, i.e., as found for the $0.0 < z < 0.25$ subsample (see Table~\ref{t_zbinFits}).
In this case, we find $C = -0.7 \pm 0.3$. We conclude that, regardless of the fitting method, the XCS-DR1 data are consistent
with negative evolution of the normalisation of the \lxt relation with respect to the self-similar expectation.

Table~\ref{t_fullFits} presents the fit parameters derived from the full sample using both the orthogonal and bisector
methods. We show results for fits with $B$ as a free parameter, and with $B$ fixed to the slope found using the
$0.0 < z < 0.25$ subsample. For ease of comparison with other works, we also list results using other common
parametrisations for the evolution of the \lxt relation in the literature. 

As noted in Section~\ref{s_intro}, while there have been several previous estimates for the evolution of the
normalisation of the \lxt relation, there is no consensus. Our result is in good agreement
with the negative evolution of the relation found by \citet[][]{Reichert_2011} 
from a heterogeneous compilation of 14 datasets, including the $z > 0.8$ \textit{XMM} Distant Cluster Project
\citep[XDCP;][]{Fassbender_2011} sample, as well as the findings of \citet{Ettori_2004} and \citet{Clerc_2012}. 
\citet{Pacaud_2007} find evolution consistent with self-similar from a sample of 24 clusters discovered in 
the \textit{XMM}-LSS survey, after correcting for selection effects, which is consistent with our result 
given the large error bar on their measurement. \citet{Maughan_2012} recently examined the \lxt relation using
a heterogeneous sample of 114 clusters drawn from the \textit{Chandra} archive, and find evolution consistent with self-similar at 
$z > 0.6$, after excising emission from cluster cores. Several other studies, based on much smaller samples, 
have found positive evolution, significantly different to that which we see here \citep[e.g.][]{Vikhlinin_2002, Lumb_2004, 
Kotov_2005}, while our result is in mild tension with the results of \citet{Novicki_2002} and 
\citet{Maughan_2006}. However, as noted by many authors, the evolution inferred is dependent upon the choice
of local \lxt relation used to set the $z = 0$ slope and normalisation used in these works. 

The main difference between the sample used here in comparison to previous works (with the exception of 
\citealt{Reichert_2011}) is the larger number of high redshift ($z > 0.6$) clusters, and it is clear from 
Fig.~\ref{f_LTNormEvo} that a long redshift baseline is needed to constrain the evolution of the relation.
It will be important to take into account both selection effects and the cluster mass function in order to 
reach a definitive conclusion. In the near future, it will be interesting to compare to measurements of the evolution of this
relation using Sunyaev-Zel'dovich effect selected cluster
samples \citep[e.g.][]{Andersson_2011}, once the number of such objects with X-ray follow-up becomes large enough.

\section{Discussion}
\label{s_discussion}

\subsection{Influence of selection effects and the cluster mass function}
\label{s_selection}

An important limitation of the analysis we have presented in this paper is that the selection function of the
survey is not taken into account. While modelling of the selection function for XCS has been performed 
\citep[see][LD11]{Sahlen_2009}, the optical follow-up required for confirmation and redshift measurements
is not complete \citep{Mehrtens_2011}, meaning that it is not currently possible to perform a more sophisticated
analysis that jointly fits for both cosmological and scaling relation parameters, while taking the selection
function into account \citep[e.g.][]{MantzScaling_2010}. The most likely selection effect that could impact our
results is Malmquist bias. For flux-limited samples, this is well known to give shallower slopes, and larger
normalisations, in scaling relations, if left uncorrected \citep*[see e.g.
Section~2.5 of the review by][]{Allen_2011}, as a consequence of objects below a luminosity threshold being 
excluded from the sample. We note that the decreasing slope with redshift seen in the fits to the XCS-DR1
sample using the bisector method (Section~\ref{s_redshiftBins}) is likely to be a manifestation of this bias.

\citet{Pacaud_2007} investigated the effect of accounting for the selection function in their measurement
of the evolution of the \lxt relation using the \textit{XMM}-LSS sample. Their sample covers a similar redshift range to
XCS-DR1 ($0.05 < z < 1.05$), but is extracted from a survey area of only 5\,deg$^2$, and so contains only
24 clusters, with 7 at $z > 0.6$. With the selection function excluded from their analysis, 
\citet{Pacaud_2007} found positive evolution of the \lxt relation with respect to self-similar ($C = 1.5 \pm 0.4$,
for a model of the form $L_{\rm X} \propto (1+z)^C$, i.e. without scaling the $L_{\rm X}$ values by $E^{-1}(z)$), 
which is significantly different to our results (see Table~\ref{t_fullFits}). This may be due to the 
different depths of the two surveys, since XCS has searched a large number of \textit{XMM} observations with longer
exposure times than \textit{XMM}-LSS (see Fig.~5 of LD11). However, after accounting for selection effects,
\citet{Pacaud_2007} find much milder evolution, which is almost exactly self-similar (although with large 
uncertainty). This demonstrates that inclusion of the selection function in the analysis acts to drive the
inferred evolution in a negative direction. Therefore it does not seem possible for uncorrected Malmquist bias
to explain the negative evolution with respect to self-similar that we see.

We have also not attempted to take into account in the analysis the effect of the (theoretically expected) 
underlying cluster distribution as a function of mass and redshift. To do this, in principle, we 
would have to assume a prior probability for the cluster temperatures and luminosities, which would be a 
decreasing function of such quantities. Given that the uncertainty in our temperature estimates tends
to increase faster with redshift than the uncertainty in our luminosity estimates, it may be that the 
end result of taking into account such an effect would be a less pronounced negative evolution of the 
normalisation of the \lxt relation. However, the size of this effect also depends on the full XCS selection
function, including follow-up incompleteness, the effect of which is likely to mitigate this bias to some extent.
We have therefore decided to defer a more detailed analysis which will quantify the size of this effect to 
future work, once we have a better understanding of the XCS follow-up incompleteness.

\subsection{Influence of cool cores, AGN, and group fraction}
\label{s_coolcores}
Another limitation, due to the serendipitous data used in this analysis, is that the low number of counts 
detected for each cluster, coupled to the low resolution off-axis and the high redshift of our sources 
(Section~\ref{s_data}), makes it unfeasible to excise the core emission from clusters, or divide the sample 
into cool-core and non-cool core populations \citep[e.g.][]{Pratt_2009, Maughan_2012}. 

This could affect our results in one of two ways. On one hand, cool core clusters are generally easier
 to detect than non-cool core clusters, due to their increased central densities. In this case, it could be 
the case that the XCS sample includes a higher fraction of cool cores than the true underlying cluster population, particularly at high
redshift. This seems not to be the case, because we see negative evolution of the \lxt relation normalisation,
and cool core clusters are known to have a higher \lxt relation normalisation than the non-cool core population 
\citep{Pratt_2009}.

On the other hand, it may be that cool core clusters at high redshift are under-represented in our sample, due
to being classified as point sources, rather than extended objects, by the detection pipeline (described in 
LD11). This could then contribute to the negative evolution of the normalisation that we see. However,
simulations in which model cool core clusters are inserted into real \textit{XMM} observations show that this is not a
significant issue for objects detected with more than 500 counts (LD11). Since the sample used in this work
contains many objects detected with $< 500$ counts, we repeated our analysis using the subsample of 108 
clusters with $> 500$ counts. We find results consistent with those found from the whole sample 
(Table~\ref{t_fullFits}), with $C = -1.6 \pm 0.6$ using the orthogonal method, and $C = -1.1 \pm 0.4$ for
the bisector method (for a model in the form of Equation~\ref{e_model}). We conclude that it is unlikely that 
a missing fraction of cool core clusters in our sample could explain our results - although it is possible that
a real evolution in the cool core fraction could cause the evolution that we see, if cool cores are less 
common at high redshift.

Similarly, although we are able to detect and excise point source emission from clusters at low redshift, this
naturally becomes increasingly difficult at high redshift, where it may be the case that contamination by AGNs is common,
as the space density of X-ray AGNs increases significantly with redshift \citep[e.g.][]{Silverman_2005}. 
Unaccounted for AGN contamination could have the effect of hardening the X-ray spectra, leading to overestimated
cluster temperatures, which could explain the negative \lxt relation normalisation evolution that we see, if 
it affects the majority of the sample. These issues can only be addressed through higher resolution X-ray imaging of the high redshift 
XCS cluster sample \citep[e.g.][]{Hilton_2010}; however, it will be possible to investigate these concerns for a small
fraction of the sample with overlapping observations in the \textit{Chandra} archive. We note that a stacking
analysis in the directions of $z > 0.9$ clusters by \citet*{Fassbender_2012} has examined this issue using
\textit{XMM-Newton} data, and finds on average one AGN within 1\,Mpc cluster-centric distance per cluster. 
While this level of contamination is thought unlikely to have a significant effect on cluster flux measurements
and sample selection, the importance of this potential bias on temperature estimates for objects in this 
redshift range has yet to be quantified.

It may be the case that there is a break in the cluster scaling laws below a certain mass or temperature threshold,
due perhaps to a change in the physics affecting the ICM between the group and cluster regimes \citep[e.g.][]{Helsdon_2000, 
Sun_2009, Stott_2012}. Similarly, \citet{Maughan_2012} see evidence for a break in the \lxt relation for high 
temperature ($T > 3.5$\,keV), relaxed systems, which seem to follow a relation consistent with the self-similar slope 
($B = 2$), whereas lower temperature, unrelaxed systems form a steeper relation. However, this effect is only
seen after the excision of core emission, which is not something that can be investigated with our data. Given
that our sample contains a number of low temperature ($T < 2$\,keV) systems, and that the fraction of low 
temperature systems decreases as redshift increases (Figs.~\ref{f_LT_sideBySide}~and~\ref{f_LT_ABC}), we 
repeat our analysis on the subsample of 149 $T > 2$\,keV clusters. Using both fitting methods, we find that 
such a cut in temperature leads to steeper slopes 
($B = 3.8 \pm 0.3$ and $B = 3.0 \pm 0.2$ for the orthogonal and bisector methods respectively), but does not
change the inferred negative evolution of the \lxt relation normalisation: we find $C = -1.8 \pm 0.6$
using the orthogonal method, and $C = -0.8 \pm 0.4$ using the bisector method, both of which are in excellent
agreement with the results obtained using the full sample of 211 clusters.

\subsection{Comparison with numerical simulations}
\label{s_sims}

Under the assumption that neither selection effects (Section~\ref{s_selection}), nor a missing cool core population or significant AGN 
contamination (Section~\ref{s_coolcores}), can explain the negative evolution of the \lxt relation normalisation that we see, we now consider the 
implications of our results for cosmological simulations of galaxy clusters. We do this by comparing to several simulations, which predict similar \lxt relations
at $z = 0$, but which behave quite differently at high redshift, as a result of the choices made in modelling the heating and 
cooling of the intracluster medium.

\subsubsection{Simulations}
Below we briefly describe the features of the models to which we compare.

\begin{table*}
\caption{\lxt relation fit parameters, derived from the bisector and orthogonal methods (see Section~\ref{s_analysis}),
for simulated cluster samples with redshift distribution matched to the XCS-DR1 spectroscopic sample used in this work.
The simulations are: CLEF \citep{Kay_2007}; the Millennium Gas simulations, for which
MG1-GO = gravity only, MG1-FO = feedback only, and MG1-PC = precooling \citep[see][]{Short_2010}; and a new version of the
Millennium Gas simulation with an updated feedback prescription (MG2-FO; \citealt*{ShortThomasYoung_2012}). The fitted model is
of the form $\log (E^{-1}(z)\,L_{\rm X}) = A + B\log (T/5) + C\log (1+z)$. The fits with $B$ fixed have the slope set to the value found from an ordinary least squares fit to the 
simulated sample at $z = 0$. Note that although the two fitting methods give significantly different values for the slope
and $z = 0$ normalisation in some cases, there is good agreement between the methods on the value of
the redshift evolution parameter, $C$.}
\label{t_simFits}
\begin{tabular}{|r|c|c|c|c|c|c|c|c|}
\hline
                    & \multicolumn{4}{|c|}{Bisector} & \multicolumn{4}{|c|}{Orthogonal}\\
Simulation & $A$ & $B$ & $C$ & $\sigma_{\log L_{\rm X}}$ & $A$ & $B$ & $C$ & $\sigma_{\log L_{\rm X}}$ \\
\hline
\multicolumn{9}{|l|}{$B$ free:}\\
CLEF & $45.31 \pm 0.04$ & $3.49 \pm 0.10$ & $-1.42 \pm 0.15$ & $0.211 \pm 0.011$ &$45.41 \pm 0.06$ & $3.99 \pm 0.19$ & $-1.25 \pm 0.24$ & $0.265 \pm 0.019$ \\
MG1-GO & $45.93 \pm 0.03$ & $2.67 \pm 0.10$ & $\phantom{+}0.16 \pm 0.10$ & $0.144 \pm 0.008$ &$46.12 \pm 0.06$ & $3.27 \pm 0.18$ & $\phantom{+}0.09 \pm 0.13$ & $0.156 \pm 0.012$ \\
MG1-FO & $44.80 \pm 0.02$ & $3.12 \pm 0.05$ & $\phantom{+}0.66 \pm 0.08$ & $0.104 \pm 0.006$ &$44.78 \pm 0.02$ & $3.23 \pm 0.07$ & $\phantom{+}0.74 \pm 0.12$ & $0.113 \pm 0.007$ \\
MG1-PC & $44.95 \pm 0.02$ & $3.92 \pm 0.06$ & $-2.20 \pm 0.06$ & $0.088 \pm 0.005$ &$44.96 \pm 0.03$ & $3.94 \pm 0.08$ & $-2.19 \pm 0.08$ & $0.088 \pm 0.006$ \\
MG2-FO & $44.44 \pm 0.02$ & $2.68 \pm 0.04$ & $\phantom{+}1.82 \pm 0.07$ & $0.094 \pm 0.006$ &$44.48 \pm 0.02$ & $2.79 \pm 0.07$ & $\phantom{+}1.79 \pm 0.10$ & $0.099 \pm 0.006$ \\

\\
\multicolumn{9}{|l|}{$B$ fixed:}\\
CLEF & $45.23 \pm 0.03$ & $3.08$ & $-1.54 \pm 0.14$ & $0.212 \pm 0.011$ &$45.26 \pm 0.04$ & $3.08$ & $-1.68 \pm 0.19$ & $0.208 \pm 0.011$ \\
MG1-GO & $45.72 \pm 0.02$ & $2.00$ & $\phantom{+}0.21 \pm 0.09$ & $0.133 \pm 0.007$ &$45.76 \pm 0.02$ & $2.00$ & $\phantom{+}0.04 \pm 0.13$ & $0.131 \pm 0.007$ \\
MG1-FO & $44.79 \pm 0.01$ & $3.30$ & $\phantom{+}0.70 \pm 0.07$ & $0.114 \pm 0.006$ &$44.77 \pm 0.02$ & $3.30$ & $\phantom{+}0.82 \pm 0.11$ & $0.113 \pm 0.007$ \\
MG1-PC & $44.76 \pm 0.01$ & $3.30$ & $-2.10 \pm 0.07$ & $0.096 \pm 0.005$ &$44.77 \pm 0.02$ & $3.30$ & $-2.13 \pm 0.10$ & $0.097 \pm 0.006$ \\
MG2-FO & $44.48 \pm 0.01$ & $2.79$ & $\phantom{+}1.74 \pm 0.07$ & $0.091 \pm 0.005$ &$44.47 \pm 0.02$ & $2.79$ & $\phantom{+}1.79 \pm 0.09$ & $0.095 \pm 0.006$ \\

\hline
\end{tabular}
\end{table*}

CLEF \citep{Kay_2007} is a hydrodynamical simulation of a 200$h^{-1}$\,Mpc 
comoving box which includes radiative cooling and feedback. The latter is implemented using the `strong feedback'
model of \citet{Kay_2004}. The amount of energy injection in this model effectively tracks the star formation
rate - a fraction of the particles which pass both a density and temperature threshold are assigned an 
entropy of 1000\,keV\,cm$^{2}$, which is then distributed through the ICM through viscous interactions and shocks.
This model produces cool core clusters at low redshift, which disappear as redshift increases, leading to a reduction
in the scatter about the relation at high redshift.

The Millennium Gas project \citep{Short_2010} is a suite of hydrodynamical simulations which
use the same volume (500$h^{-1}$\,Mpc$^{3}$) and initial perturbations as the Millennium Simulation \citep{Springel_2005}.
This set of simulations includes a gravity only `control' model (MG1-GO); a simulation with energy injection at high redshift
and radiative cooling (we refer to this as the `precooling' model, or MG1-PC); and a simulation which incorporates
feedback from AGN and supernovae, implemented using a semi-analytic model (MG1-FO). 

The MG1-PC simulation implements preheating of the cluster gas at high redshift in a similar
fashion to previous work \citep[e.g.][]{Bialek_2001, Borgani_2002}. In this case, the entropy
of each gas particle is raised to 200\,keV\,cm$^2$ at $z=4$. While this model is not physically plausible
(only two per cent of the baryons form stars by $z=0$, and the model is incapable of forming cool core clusters; 
\citealt{Short_2010}), it does reproduce the \lxt relation at $z=0$ \citep{Hartley_2008}.

The MG1-FO model includes supernova and AGN feedback using the scheme of \citet{ShortThomas_2009},
where the semi-analytic galaxy formation model employed by \citet{DeLucia_2007} is used to infer
both the star formation rate (a sink of hot gas) and the heating rate due to supernovae and AGN. The AGN
feedback is implemented using the scheme suggested by \citet{Bower_2008} and is capped at two per cent
of the Eddington rate (see \citealt{Short_2010} for full details). The model successfully reproduces both the
local \lxt relation \citep{Short_2010}, and the Sunyaev-Zel'dovich $Y-M$ relation \citep{Kay_2012}. 

We also compare to an updated version of the Millennium Gas model with AGN feedback (MG2-FO). This
run was performed using a 250$h^{-1}$\,Mpc box with higher resolution and updated cosmological parameters 
(consistent with the WMAP 7-year results; \citealt{Komatsu_2011}). The semi-analytic galaxy formation model is
also newer \citep{Guo_2011}, and the feedback is now implemented in a stochastic fashion (only a fraction of
the intracluster gas particles are heated directly, whereas in the previous model the energy was shared throughout
the cluster). The model improves agreement with non cool-core clusters but, like the previous Millennium Gas
models, fails to produce the cool-core population due to the absence of radiative cooling. Full details of this 
implementation are discussed in \citet*{ShortThomasYoung_2012}.

\subsubsection{Results}
All of the simulations described above provide measurements that are comparable to the real XCS-DR1 data. In all
cases, we use total (i.e. core included) bolometric luminosity measurements within $R_{500}$, which is defined
with respect to the critical density, as for the XCS-DR1 measurements. We note that each of these simulations
assumes a slightly different cosmology to the one assumed in this work. We have neglected to correct 
the luminosities to account for this, as it is a small effect, and does not significantly affect the evolution
of the normalisation of the relation, which differs substantially between the models, due to the different
physical assumptions in each. 

\begin{figure*}
\includegraphics[width=14 cm]{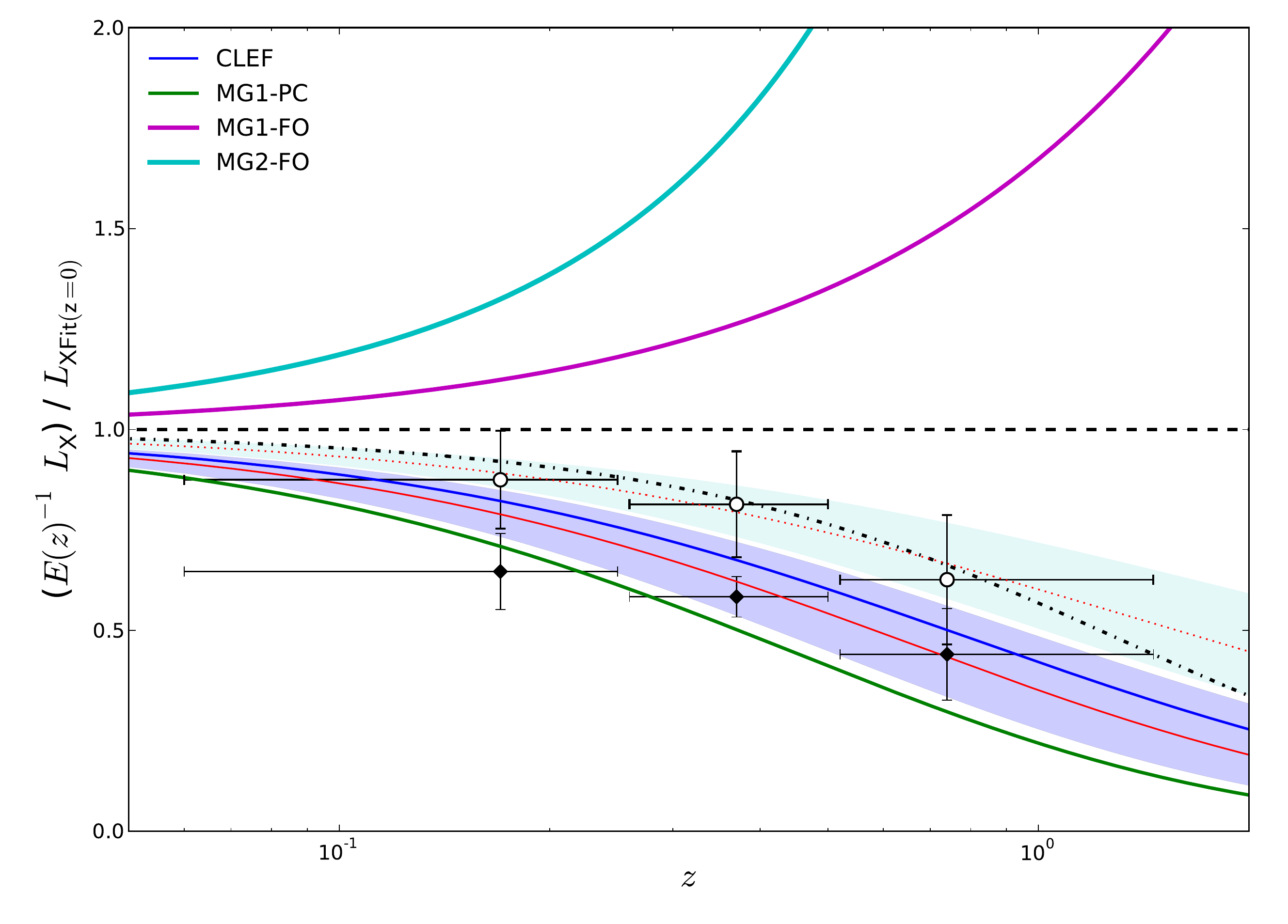}
\caption{Evolution of the normalisation of the \lxt relation as measured from XCS-DR1, compared with numerical
simulations (indicated in the legend; see Section~\ref{s_sims} and Table~\ref{t_simFits} for details). The
solid thin red line is the best-fit obtained from XCS-DR1 using the orthogonal method, while the black diamonds show the
corresponding median values for the clusters in each redshift bin (horizontal error bars indicate the redshift
range of each bin, while vertical error bars indicate the 1$\sigma$ uncertainty in the median, estimated using
bootstrap resampling). The dotted thin red line similarly represents the 
best-fit for XCS-DR1 obtained using the bisector method, with the slope fixed at the $z = 0$ value (2.81, see 
Table~\ref{t_fullFits}), and the open circles indicate the median values for clusters in 
redshift bins for this set of best-fit \lxt relation parameters. This latter fit is consistent with no 
evolution (dot-dashed line). The shaded regions mark the marginalised
68 per cent confidence regions. While the amount of evolution inferred from the XCS sample depends on the 
fitting technique used, there is no such dependence for the simulated data (see Table~\ref{t_simFits}), 
and so we only show the results of the fits to the simulations using the orthogonal method.
The XCS data favour negative evolution with respect to self-similar ($E(z)$; horizontal dotted line), and
are clearly better described by the CLEF or MG1-PC models, rather than
the models which implement AGN and supernovae feedback using a semi-analytic prescription (MG1-FO and MG2-FO).}
\label{f_simComparison}
\end{figure*}

All of the simulations
provide spectroscopic-like temperature estimates \citep[$T_{\rm sl}$;][]{Mazzotta_2004}. We restrict our analysis
in all cases to clusters with $T_{\rm sl} > 2$\,keV, as this is the regime in which the bremsstrahlung mechanism
dominates, and is where the spectroscopic-like temperatures are most robust. Objects identified as satellites
to more massive haloes are not included in the samples we draw from the simulations.

For consistency with the analysis of the XCS data, we draw random samples from each simulation with redshift
distributions matched to that of XCS-DR1 (Fig.~\ref{f_zT_histograms}), and we fit each sample using both 
the orthogonal and bisector methods, as before. Table~\ref{t_simFits} lists the results. As is the case for the real data, for some 
simulations (e.g. CLEF), the bisector and orthogonal methods give different slopes and normalisations, with
the orthogonal slope being steeper. However, we see that in all cases, both methods give consistent values for
the redshift evolution parameter, $C$. The better agreement in $C$ between the two fitting methods, when used
on the simulations as compared to the real data, is likely to be due to the absence of selection effects in 
the former.

In some cases, we find steeper slopes than were measured in the original works describing the simulations, 
most notably in the MG1-GO case, where $B=2$ is expected. This is due to our fitting methods (both \citealt{Kay_2007} and \citealt{Short_2010}
use ordinary least squares regression). We checked that this does not bias our estimates of $C$, by repeating the 
fitting with the slope ($B$) fixed to the values found in \citet{Kay_2007}
and \citet{Short_2010} from the complete simulated samples. These results are also listed in Table~\ref{t_simFits}.
In all cases, when $B$ is fixed, the values of $C$ change by at 
most 2$\sigma$ in comparison to the fits with $B$ as a free parameter. In no case do we find qualitatively different
behaviour for a given model as a result of changing the fitting technique or fixing the slope: e.g., we find negative
evolution of the \lxt relation in CLEF for all the variations.

Fig.~\ref{f_simComparison} shows a comparison of the redshift evolution in the simulations with the results
from the XCS-DR1 sample, where we show the results for both the orthogonal fitting method, and the bisector
method with slope fixed to the $z = 0$ value (see Section~\ref{s_redshiftBins} and Table~\ref{t_fullFits}).
This gives an indication of the systematic uncertainty in the XCS-DR1 measurement due to the choice of the fitting method. We see
that in either case the XCS-DR1 data are closer to the CLEF and MG1-PC simulations, in which the \lxt 
normalisation evolves negatively with respect to self-similar, and are more than $5\sigma$ away
from the evolution predicted in the MG1-FO and MG2-FO simulations (irrespective of the fitting
method used on the XCS-DR1 data).

The key difference between the feedback models in the simulation is the epoch at which most of the energy
injection occurs. In the MG1-PC model, all of the energy input occurs at $z=4$, which is not likely to be physically
reasonable, but serves as a useful extreme test. In the CLEF simulation, the energy injection occurs over a broad
range of redshifts, but is skewed to early times, as it directly tracks the star formation rate (around two thirds of the 
stars have already formed, and energy injected, by $z=1$). Finally, in the MG1-FO and MG2-FO simulations, the dominant
AGN feedback occurs later, when the black holes have grown to sufficient mass to act as powerful energy sources.
Therefore, the lack of agreement with the observations suggests that feedback at high redshift is too inefficient in the
current models. We note also that radiative cooling is not currently implemented in these simulations, and therefore
the cold gas mass growth rate in the semi-analytic model is not fully self consistent with the hydrodynamical simulation.


\section{Conclusions}
\label{s_conclusions}

We have investigated the evolution of the \lxt relation since $z \sim 1.5$ using a sample of 211
spectroscopically confirmed X-ray clusters drawn from the first XCS data release \citep{Mehrtens_2011}. This
is the first such measurement over this wide redshift range using a single, homogeneous sample. We 
find:

\begin{enumerate}

\item{Using both an orthogonal and bisector fitting method, the slope of the \lxt relation for the $z < 0.25$
subsample of XCS-DR1 clusters is consistent with that found for the REXCESS sample \citep{Pratt_2009}. The 
normalisation is slightly lower, but consistent within $2\sigma$, using the orthogonal method, although we find
a 5$\sigma$ lower normalisation using the bisector method. This may be in part due to differences in the 
spectral fitting, or could be due to differences in the sample selection.}

\item{From dividing the sample into redshift bins, using the orthogonal fitting method, we see no evidence
for evolution in either the slope or intrinsic scatter as redshift increases - both are consistent with 
previous measurements at $z = 0.1$. We see a flattening of the slope at high redshift when using the bisector
fitting method, which could be a signature of the effect of Malmquist bias.}
\vskip 3pt

\item{Regardless of the fitting method, our data shows that the normalisation of the relation evolves negatively with
respect to self-similar. For the orthogonal method, we find the evolution is 
$E^{-1}(z)\,L_{\rm X} = 10^{44.67 \pm 0.09} (T/5)^{3.04 \pm 0.16} (1+z)^{-1.5 \pm 0.5}$,
which is within $2\sigma$ of the zero evolution case. Using the bisector method, with the slope fixed to the value
found for the $z < 0.25$ subsample, we find $E^{-1}(z)\,L_{\rm X} \propto T^{2.81}(1+z)^{-0.7 \pm 0.3}$. Malmquist 
bias would have the effect of driving the normalisation in the positive direction, and so cannot explain this
result. It is possible that a deficit of cool cores in the XCS-DR1 sample, or significant AGN contamination
at high redshift, may contribute to the negative evolution that we see. The former seems 
unlikely, given that a higher signal-to-noise subsample gives consistent results to those obtained
using the full sample, while the latter can only be tested using higher resolution X-ray data.}
\vskip 3pt

\item{From comparison with numerical simulations, we find the XCS-DR1 data favour feedback models in which the
majority of the energy injection occurs at high redshift. AGN feedback models based on current semi-analytic
galaxy formation model prescriptions, as used in the Millennium Gas project, predict positive evolution with 
respect to self-similar, and differ from the XCS-DR1 measurements at the $>5 \sigma$ level. This suggests
that feedback at high redshift in these models is too inefficient.}
\vskip 3pt
\end{enumerate}

A more sophisticated analysis to jointly constrain both cosmological and scaling relation parameters, taking into
account both a model of the survey selection function and the cluster mass function, will be possible with 
improved redshift completeness. We are
also pursuing velocity dispersion measurements of the high redshift XCS cluster sample, and will explore 
the evolution of the scaling of X-ray observables with dynamical mass in future work.

\section*{Acknowledgments}

We thank the referee for a thoughtful report which improved the clarity of this paper. We thank Eric Miller 
and Gabriel Pratt for useful discussions. Financial support for this project was provided 
by: the Science and Technology Facilities Council (STFC) through grants ST/F002858/1 and/or ST/I000976/1 
(for EL-D, AKR, NM, MHo, ARL and MS), ST/H002391/1 and PP/E001149/1 (for CAC), ST/G002592/1 (for STK); 
the Leverhulme Trust (for MHi); the University of KwaZulu-Natal (for MHi); the University of Sussex (for MHo);
FP7-PEOPLE-2007-43-IRG n 20218 (for BH); Funda\c{c}ao para a Ci\^{e}ncia e a Tecnologia through the project 
PTDC/CTE-AST/64711/2006 (for PTPV); the South East Physics Network (for RCN); the Swedish Research Council 
(VR) through the Oskar Klein Centre for Cosmoparticle Physics (for MS); the RAS Hosie Bequest and the 
University of Edinburgh (for MD); the U.S. Department of Energy, National Nuclear Security Administration by 
the University of California, Lawrence Livermore National Laboratory under contract No. W-7405-Eng-48 
(for SAS). JPS acknowledges support from STFC. ARL was supported by a Royal Society--Wolfson Research Merit 
Award.


\bibliographystyle{mn2e}
\bibliography{refs}

\label{lastpage}

\end{document}